\documentstyle[preprint,aps,prabib]{revtex} 

\begin{document}

\title{Gravitational Flux Tubes}

\author{V. Dzhunushaliev \thanks{E-mail:
dzhun@freenet.bishkek.su
and dzhun@rz.uni-potsdam.de}}
\address{Universit\"at Potsdam, Institut f\"ur Mathematik,
14469, Potsdam, Germany, and
Theor. Phys. Dept. KSNU, 720024, Bishkek, Kyrgyzstan}

\author{U. Kasper \thanks{Trojanstr. 7, D-12437 Berlin,
Email: ukasper@rz.uni-potsdam.de}}

\author{D. Singleton \thanks{E-mail:
das3y@zimmer.csufresno.edu}}
\address{Dept. of Phys. CSU Fresno, 2345 East San Ramon Ave.
M/S 37 Fresno, CA 93740-8031, USA}

\date{\today}

\maketitle

\begin{abstract}

By studying multidimensional Kaluza-Klein theories,
or gravity plus U(1) or SU(2) gauge fields it is
shown that these theories possess similar flux
tube solutions. The gauge field which fills the tube
geometry of these solutions leads to a comparision with the
flux tube structures in QCD. These solutions also carry
a ``magnetic'' charge, $Q$, which for the SU(2)
Einstein-Yang-Mills (EYM) system exhibits a dual relationship
with the Yang-Mills gauge coupling, $g$, ($Q=1/g$). As $Q \to 0$
or $Q \to \infty$, $g \to \infty$ or $g \to 0$ respectively.
Thus within this classical EYM field theory
we find solutions which have features --
flux tubes, magnetic charges, large
value of the gauge coupling -- that are
similar to the key ingredients of confinement in QCD.
\end{abstract}

\narrowtext

\section{Introduction}

In this article we consider space-times of dimension
$D=$ 4, 5, or 7. In the $D=4$ case
we also include gauge fields, while in the
higher dimensional cases the gauge fields arise
naturally from the extra dimensions in the Kaluza-Klein
way. The external 4-dimensional (4D) space-times that we
consider have the topological
structure ${\bf R^1}\times{\bf R^1}\times{\bf D^2}$ or
${\bf R^1}\times{\bf R^1}\times{\bf S^2}$ (where
${\bf D^2}$ is topologically the 2-dimensional (2D)
disk and ${\bf S^2}$ the 2D sphere). In both
cases we take the external 3-dimensional (3D)
spaces to have a tube structure so that locally the
metric appears as
\begin{equation}
\label{1}
ds_3^2=dr^2+a^2(d\theta^2+\sin ^2(\theta) d\phi^2)
\end{equation}
with {\it a} being a constant. For the first topology the
coordinates take the range
\begin{equation}
\label{2}
-\infty< r +<\infty\quad ,\; \; \; \; \;
0< \theta < \pi\quad ,\; \; \; \; \;
-\pi/2< \phi < \pi/2
\end{equation}
while in the second case the $\phi$ range is altered to
\begin{equation}
\label{2a}
-\pi< \phi <\pi
\end{equation}
These ranges are in accordance with the usual conventions
for the 2D disk and 2D sphere respectively. These two cases
differ in the identification of the boundary points.
\par
The metric of the external 4D space-time is then
\begin{equation}
\label{3}
ds^2=e^{2\nu(r)}dt^2- ds_3 ^2\quad .
\end{equation}
This type of metric was considered in Ref.
\cite{Levi-Civita} with the source term in Einstein's equations
being  an electric field. The ansatz function, $e^{\nu}$, for
the flux tube solution of this 4D Einstein-Maxwell system is
given by
\begin{equation}
\label{4}
e^{\nu}=c_1e^{r/a}+c_2e^{-r/a}
\end{equation}
with $c_1$, $c_2$ being constants of integration.
\par
In Ref. \cite{Bertotti} a magnetic field
was included in addition to the electric field.
The form of these fields was given by
\begin{equation}
\label{5}
F_{10}= \rho ^{1/2} \cos \alpha e^{\nu} \; \; \; \; \;
F_{23} = a^2 \rho ^{1/2} \sin \alpha \sin ^2 (\theta)
\end{equation}
where $\alpha$ is an arbitrary constant and $\rho$ satisfies
\begin{equation}
\label{6}
\rho = \frac{8 \pi}{\kappa a^2} \quad .
\end{equation}
$\kappa$ is Einstein's constant. We use units such that
$c=1$ and $\hbar =1$ so that $\kappa$ has dimension $length^2$
and the electromagnetic charges are dimensionless. In
this case $c_1, c_2$ are required to take values such that
$e^{\nu(r)} = \cosh (r/a)$ in Eq. (\ref{4}). With this
the metric in Eq.(\ref{3}) becomes
\begin{equation}
\label{7}
ds^2=\cosh^2(r/a)dt^2-
dr^2-a^2(d\theta^2+\sin^2(\theta)d\phi^2)
\end{equation}
It is possible to chose coordinates such that $F_{01}$ and
$F_{23}$ are constant. In Ref. \cite{Guendelman}, a metric of this
type was considered for the construction of compactified phases in 4D
space-times\footnote{We note that the components of
the electromagnetic field (2b) in \cite{Guendelman} are not
those belonging to the coordinate system chosen in (2a).)}.
\par
Having briefly introduced the flux tube solution for
the 4D Einstein-Maxwell system, we will turn in the following
sections to examining 5D and 7D Kaluza-Klein metrics, and
4D Einstein-Yang-Mills systems all of which exhibit similar
flux tube solutions ({\it i.e.} metrics whose 3D space
is like that in Eq. (\ref{1}), and with gauge fields
running down the length of the tube structure). We
will also point out some important differences between
these solution ({\it e.g.} the 4D space-time of the
7D solution with the internal group of SU(2) {\it is not}
a solution of the 4D Einstein-Yang-Mills equations
with gauge group SU(2)).
\par
The ``flux tube'' solutions that are presented here bear
some resemblance to the flux tubes which occur
in type-II superconductors and also in quantum
chromodynamics (QCD). In QCD a tube of
constant chromodynamic fields is thought to form between
quark-antiquark pairs. In the gravitational solutions
presented the 3D hypersurfaces, $t=const$, are
tubes filled with some gauge field. The flux tubes
in QCD are thought to be closely connected with the confinement
mechanism. The similarity between the QCD flux tubes and
the gravitational flux tubes might indicate some
analogous {\it classical} confinement mechanism for
gravity theories ({\it i.e.} these solution might
indicate a gravitational ``confinement'' of two (chromo)electrical
and/or (chromo) magnetic charges via a flux tube stretched between
them).
\par
The main goals of our paper are to make the case :
\begin{itemize}
\item
that the flux tube solutions are typical solutions of gravitational
theories in the presence of gauge fields.
\item
that these gravitational solutions may indicate some link
with the flux tubes of QCD. Further, for the 4D Einstein-Yang-Mills
(EYM) system we find a very strong similarity between the
structure of the solutions and some key features of the
standard mechanism of confinement of color charge.
\end{itemize}

\section{5D Kaluza-Klein Theory}

In this section we will look at a 5D Kaluza-Klein version
of the 4D solution given in the introduction, and compare
the two cases. The general 5D Kaluza-Klein metric is given by
\begin{equation}
\label{9}
ds^2_5=g_{\mu\nu}dx^{\mu}dx^{\nu}-
J\left(r_0 d\chi+\sqrt{\kappa/4\pi}
A_{\mu}dx^{\mu}\right)^2
\end{equation}
with the 5$^{th}$ coordinate, $\chi$, being restricted by
$0<\chi<2\pi $, and the end points of this interval are
identified. $A_{\mu}$ is the electromagnetic potential,
and $r_0$ is an arbitrary constant. In the spherically
symmetric case this give the following 5D metric
\begin{eqnarray}
\label{10}
ds_5^2&=&e^{2\nu(r)}dt^2-dr^2-a^2(d\theta^2+\sin^2(\theta)d\phi^2)\nonumber\\
&&-\left(r_0 d\chi +\sqrt{\frac{\kappa}{4\pi}}\omega(r)dt-
\sqrt{\frac{\kappa}{4\pi}} Q \cos(\theta)d\phi\right)^2
\end{eqnarray}
with $J =1$ and, as before, $a=const$.
The non-vanishing components of the
electromagnetic potential are
\begin{equation}
\label{11}
A_0(r)=\omega(r) \; \; \; \; \; \; ,
A_3(\theta)=-\frac{Q}{a}\cos(\theta)\quad .
\end{equation}
The $A_3$ component is formally that of a magnetic monopole. With
this ansatz, the 5D Kaluza-Klein field equations become
\cite{Dzhunu}
\begin{eqnarray}
\label{14}
\omega''  &-& \nu ' \omega ' =  0 \quad , \; \; \; \; \; \;
\; \; \; \; \; \; \; \; \; \; \; \; \; \; \; \; \; \; \; \; \;
\frac{1}{a^2}  =  \frac{\kappa}{8\pi}\frac{Q^2}{a^4}\quad
\\ \nonumber
\nu'' &+&  (\nu')^2-\frac{\kappa}{8 \pi}(\omega ')^2 e^{-2 \nu}=0\quad ,
\; \; \; \; \;
a^4(\omega')^2-e^{2\nu}Q^2=0\quad .
\end{eqnarray}
The first integral to the first equation of (\ref{14})
is given by
\begin{equation}
\label{17}
\omega ' = \frac{q}{a^2} e^{\nu} \quad .
\end{equation}
$q$ is a constant of integration. With Eq. (\ref{17}) the
last equation of (\ref{14}) reads
\begin{equation}
\label{17a}
q^2=Q^2
\end{equation}
or, equivalently,
\begin{equation}
\label{aaa}
\frac{\kappa}{16\pi}(Q^2+q^2)=a^2 \quad .
\end{equation}
Finally
one can use $\omega '$ in the third equation of
Eq. (\ref{14}) and also integrate it directly to
obtain
\begin{equation}
\label{17aa}
        e^{\nu}=\cosh \left( {\frac{r}{a}} \right)\quad , \; \; \;
\; \; \; \; \;
\omega=\frac{q}{a} \sinh \left( \frac{r}{a} \right)
\quad .
\end{equation}
In some limited sense $Q$ can be interpreted as a magnetic
charge, and $q$ as an electric charge. Both $q$ and $Q$ appear
formally in the expressions of $A_0(r)$ and $A_3(\theta)$
in the same way in which charges usually appear. Further,
as $q$ and $Q$ increase the strength of the electromagnetic
field increases in a directly proportional manner. However,
strictly there are no charges present since we are dealing
with the vacuum equations, and there are no singularities in the
electromagnetic potential.
\par
This 5D solution is nearly identical to the 4D solution studied
in the previous section, except in the present case we have the
additional restriction $q^2=Q^2$, which fixes the strengths of
the ``electric'' and ``magnetic'' field to be equal as we will show next.
In the 4D case one could vary the relative strengths of the
electromagnetic fields (by varying $\alpha$).
This equality between ``electric'' and
``magnetic'' fields was already noticed for other
Kaluza-Klein dyon solutions \cite{Perry}.
\par
As in Ref. \cite{Landau} we now define the two pairs of
electromagnetic 3-vectors $\bf E, B$
and $\bf D, H$
\begin{equation}
\label{17b}
E_{\alpha}=F_{0\alpha}\quad ,  \; \; \; \; \;
B^{\alpha}=-\frac{1}{2\sqrt{\gamma}}\epsilon^{\alpha \beta
\rho}F_{\beta \rho}\quad ,
\end{equation}
and
\begin{equation}
\label{18}
D^{\alpha}=-\sqrt{g_{00}}F^{0\alpha}\quad , \; \; \; \; \;
H_{\alpha}=-\frac{1}{2}\sqrt{\gamma}\epsilon_{\alpha \beta
\mu}\sqrt{g_{00}}F^{\beta \mu}\quad
\end{equation}
with $\gamma$ the determinant of the 3D metric
of the $t=const$ hypersurfaces, and
$\epsilon_{\alpha\beta\gamma}$ is the
Levi-Civita symbol.
The only non-vanishing components are
\begin{equation}
\label{19a}
E_1=-\frac{q}{a^2}e^{\nu}\quad \; \; \; \; \;
B^1=-\frac{q}{a^2}\quad ,
\end{equation}
and
\begin{equation}
\label{20}
D^1=-\frac{q}{a^2}\quad , \; \; \; \; \;
H_1=- \frac{q}{a^2}\quad .
\end{equation}
The 3D electromagnetic field vectors $\bf D, H, B$ as defined
above, are constant but $\bf E$ is not.

\section{Einstein-Yang-Mills Theory}

In this section, we will examine gravity plus an SU(2) Yang-Mills
field. The 4D metric that we consider is identical to that given in
Eq. (\ref{3}).
\par
For the gauge potential components $A_{\mu}^a$
we use the ansatz (see also Refs. \cite{Wu-Yang},
\cite{'t Hooft}) in spherical-polar coordinates
\begin{eqnarray}
\label{22}
\left(A^a_{\theta}\right)&=&\frac{Q}{a}(\sin\phi,-\cos\phi,0)\quad
, \nonumber \\
\left(A^a_{\phi}\right)&=&\frac{Q}{a}\sin\theta(\cos\phi\cos\theta,
\sin\phi\cos\theta, -\sin\theta)\quad ,\\
\left(A^a_{t}\right)&=&V(r)(\sin\theta\cos\phi,
\sin\theta\sin\phi, \cos\theta)\quad \nonumber.
\end{eqnarray}
All other components are zero. The field
strength components are defined via
\begin{equation}
\label{22a}
F^a_{\mu\nu}=A^a_{\nu,\mu}-A^a_{\mu,\nu}
+g\epsilon^{abc}A^b_{\mu}A^c_{\nu}
\end{equation}
with $g$ being the Yang-Mills coupling constant.
Under these assumptions the gravitational part of the
EYM equations reduce to
\begin{equation}
\label{aa}
\frac{1}{a^2}=\frac{\kappa e^{-2\nu}}{8\pi
a^4}\left[V'^2a^4+e^{2\nu}Q^2(gQ-2)^2-2a^2V^2(gQ-1)^2
\right]\quad ,
\end{equation}
\begin{equation}
\label{bb}
a^2(\nu''+\nu'^2)=\frac{\kappa e^{-2\nu}}{8\pi
a^2}\left[V'^2a^4+e^{2\nu}Q^2(gQ-2)^2\right]\quad ,
\end{equation}
\begin{equation}
\label{cc}
\frac{1}{a^2}=\frac{\kappa e^{-2\nu}}{8\pi
a^4}\left[V'^2a^4+e^{2\nu}Q^2(gQ-2)^2+2a^2V^2(gQ-1)^2
\right]\quad .
\end{equation}
The difference of (\ref{aa}) and (\ref{cc}) leads to
\begin{equation}
\label{cc1}
Qg-1=0 \rightarrow Q=\frac{1}{g}.
\end{equation}
Taking into account this dual restriction between $Q$ and
$g$ the only non zero equation for the gauge field
part of the EYM field equations is
\begin{equation}
\label{dd}
V''-\nu'V'=0\quad .
\end{equation}
This equation has the first integral
\begin{equation}
\label{dd2}
V'=\frac{q}{a^2}e^{\nu}
\end{equation}
with $q$ being a constant of integration.
The third of the EYM equations, Eq. (\ref{cc}), together
with Eq. (\ref{dd2}) and the duality condition between $Q$ and $g$
leads to the following relation between parameters of the model and the
integration constant
\begin{equation}
\label{cc3}
\frac{\kappa}{8\pi}(q^2+Q^2)=a^2\quad .
\end{equation}
This expression is similar to (\ref{17a}) we obtained for
the model in the 5D Kaluza-Klein theory. Finally, a special
solution of Eq. (\ref{bb}) is, as in the 5D Kaluza-Klein theory,
\begin{equation}
\label{cc2}
e^{\nu(r)}=\cosh\left(\frac{r}{a}\right)\quad .
\end{equation}
Using this expression for $e^{\nu (r)}$ it is possible to
integrate $V'$ of Eq. (\ref{dd2}) to obtain $V (r)$ in a form
similar to $\omega (r)$ of Eq. (\ref{17aa}).
There is an important difference between the present solution
and the previous Abelian one: In the latter case the
integration constant $q$ has to be equal to the parameter
$Q$ and $Q$ by itself is given in terms of $\kappa$ and $a^2$.
\par
Based on the above solution one can put forth a rough picture
of the confinement mechanism which involves the presence of
gravity in a crucial way, rather than confinement being solely due to
the Yang-Mills fields. At large energy scales it is more
likely for ``magnetic'' charges to fluctuate out of the
vacuum. If magnetic charge is quantized by some fundamental
quanta, $Q_0$, then $Q = n Q_0$ where $n = 0, 1, 2, 3 ...$.
At larger energy scales it becomes more likely to have larger
$n$ and therefore larger $Q$ fluctuate out of the vacuum. The
higher $n$ values can be viewed either as $n$ magnetic
charges of unit $Q_0$, or a single magnetic charge which
carries a multiple, $n$, of the fundamental magnetic charge, $Q_0$.
Thus as the energy scale increases $Q$ increases and $g$
decreases taking on small, perturbative values. 
As one lowers the energy scale the fluctuation of ``magnetic''
charges out of the vacuum becomes less probable so that
$Q \rightarrow 0$ and $g \rightarrow \infty$. Thus in the
high and low energy scale limits this EYM system
can be seen as having two differing phases : For
$Q$ finite and large, $g$ is small and perturbative, and
one has a deconfined phase; for $Q \to 0$, $g$ is large
and non-perturbative and
one has a confined phase. Many features of the standard
picture of confinement (flux tube structures, ``magnetic'' charges,
and a gauge coupling $g$ which can become small or large)
are embodied by these solutions. Unlike the
standard picture where one has only Yang-Mills fields, for the
present solutions the presence of the gravitational interaction
appears to play a significant role.
\par
Next we look at the ``electromagnetic'' field embedded in
the non-Abelian gauge fields. As in Refs. \cite{'t Hooft} ,
\cite{Arafune}, we define the components of the
``electromagnetic'' U(1) gauge potential by
\begin{equation}
\label{28}
A_{\mu}=n_aA^a_{\mu}\quad .
\end{equation}
where $(n_1, n_2, n_3) = (\sin{\theta}\cos{\phi},
\sin{\theta}\sin{\phi},\cos{\theta})$ is the normal
vector to the unit sphere. For the ansatz
given in Eq. (\ref{22}) this yields
\begin{equation}
\label{29}
A_{\mu}=V\delta^0_{\mu}\quad .
\end{equation}
One possible definition of an electromagnetic field
tensor, which is invariant with respect to SU(2)
transformations is
\begin{equation}
\label{30}
F_{\mu\nu}=A_{\nu,\mu}-A_{\nu,\mu}
-\epsilon_{abc}n^an^b_{,\mu}n^c_{,\nu}\quad.
\end{equation}
For the present ansatz this leads to
\begin{equation}
\label{31}
F_{10} = V'\quad , \; \; \; \; \; \; F_{23} = -\frac{Q}{a}\sin{\theta}\quad .
\end{equation}
The electromagnetic field defined this way is a solution of
the 4D Einstein-Maxwell {\it vacuum} field equations. Although this form
of the gauge potentials and field strength tensor is
reminiscent of a magnetic monopole it possesses an additional
electric component and its topological structure does
not allow the localization of a source.

\section{Special vacuum equations for D=7}

As in Ref.\cite{Dzhunu2} the multidimensional gravity
theories considered in this paper are gravity
on the principal bundle with either U(1) or SU(2) as
the structural group. The easiest way to derive the field
equations in the last case is to change over to the 7-bein formalism
(with $h^{\bar{A}}_B$ the components of the 7-bein).
The metric in this case is
\begin{equation}
ds^2 = \omega ^{\bar{\mu}} \omega _{\bar{\mu}} +
\omega ^{\bar{a}} \omega _{\bar{a}}
\label{34c}
\end{equation}
with
\begin{equation}
\omega ^{\bar{\mu}} = h^{\bar{\mu}}_\nu dx^\nu,
\; \; \; \; \; \;
\omega ^{\bar{a}} = \varphi(x^\mu)\left(r_0 \sigma ^{\bar{a}} +
A^{\bar{a}}_\nu (x^\mu)dx^\nu\right)\quad  .
\label{34b}
\end{equation}
An overbar means a component with respect to an
orthonormalized basis.
Here, $\sigma ^{\bar{a}} = h^{\bar{a}}_b dx ^b$ are the
Maurer-Cartan 1-forms on the group, and $A^{\bar{a}}_\nu$ is
the gauge potential. Capital letters run from 0 to 6,
lower case Greek letters from 0 to 3, and lower case Latin
letters from 4 to 6. We set $h^{\bar{\mu}}_b=0$ and do not
vary them, so that we have only the following
independent degrees of freedom:
$h^{\bar{A}}_\nu = \left \{h^{\bar{\mu}}_\nu ;
A^{\bar{a}}_\nu\right \}$ and $\varphi(x^\mu)$. Since the fiber
of the principal bundle is a symmetric space (the SU(2) gauge
group in this section) $h^{\bar{a}}_b$ is given and can not be
varied (for the SU(2) gauge group they will given below).
For the Lagrange density we choose the 7D curvature scalar
density $h\tilde{R}$ ($h=det({h^{\bar{A}}_B})$; $\tilde{R}$
is the Ricci scalar).
Varying with respect to only the independent degrees of freedom
leads to the following gravitational field equations
\begin{equation}
\label{36}
\tilde{R}_{\bar{\mu}}^A-\frac{1}{2}h_{\bar{\mu}}^A\tilde{R}=0
\end{equation}
and
\begin{equation}
\label{37}
\left(\tilde{R}_{\bar{a}}^b-\frac{1}{2}h_{\bar{a}}^b\tilde{R}\right)
h^{\bar{a}}_b=0
\end{equation}
Eqs. (\ref{36}), with indices $\left ({a}\atop{\bar{\mu}}\right )$,
are the ``Yang-Mills'' equations.
\par
Now, we consider a 7D space-time with a metric which is
similar to the metric of Eq. (\ref{10})
(with $\varphi (x ^{\mu}) =1$ in Eq. (\ref{34b})) \cite{Dzhunu}
\begin{equation}
\label{34}
ds^2_7=e^{2\nu(r)}dt^2-dr^2-a^2\left(d\theta^2+\sin^2{\theta}
d\phi^2\right)-
\sum_{a=1}^3 \left(r_0\sigma^a-\sqrt{\frac{\kappa}{4\pi}}
A^{\bar a}_{\mu}(r)dx^{\mu}\right)^2 .
\end{equation}
This form of the metric is also similar to the metric
of certain multidimensional cosmology models in that the metric
coefficients of the internal metric depend only on the coordinates of
the external 4D space-time. The length
of the internal space is characterized by $r_0$.
\par
Again, $a$ is a non-zero constant. The metric of
the external 4D space-time is given by the first three
terms in (\ref{34}).
The three 1-forms - $\sigma ^a$ ($a=1, 2, 3$) -
are, by definition, related to the Euler angles
$\alpha$, $\beta$, $\gamma$ of the $SU(2)$ group via
\begin{eqnarray}
\label{23}
\sigma^1&=&\left(\sin{\alpha}
d\beta-\sin{\beta}\cos{\alpha}d\gamma\right)\quad ,\\
\sigma^2&=&-\left(\cos{\alpha}
d\beta+\sin{\beta}\sin{\alpha}d\gamma\right)\quad ,\\
\sigma^3&=&\left(d\alpha+\cos{\beta}
d\gamma\right)\quad .
\end{eqnarray}
The ansatz for $A_{\mu}^{\bar a}$ is similar to Eq. (\ref{22})
\begin{eqnarray}
\label{35}
\left(A^{\bar a}_{\theta}\right)&=&\frac{Q}{a}\left(\sin{\phi},
-\cos{\phi},0\right)\quad
, \nonumber \\
\left(A^{\bar a}_{\phi}\right)&=&\frac{Q}{a}\sin{\theta}
\left(\cos{\phi}\cos{\theta},
\sin{\phi}\cos{\theta},
-\sin{\theta}\right) \\
\left(A^{\bar a} _t\right)&=&V(r)\left(\sin{\theta}
\cos{\phi},
\sin{\theta}\sin{\phi},
\cos{\theta}\right)\nonumber\quad .
\end{eqnarray}
All other components are zero.
The remarks made in connection with the physical and
topological meaning of this ansatz are similar to
the remarks made about the ansatz for the Einstein-Yang-Mills
system in the previous section.

Now, let us write down the 4D Einstein-type equations following
from $\tilde{R}_1^1=0$
and $\tilde{R}_2^2=0$. They read
\begin{equation}
\label{aaa1}
\nu''+(\nu')^2=\frac{\kappa}{8\pi}e^{- 2 \nu}V'^2
-\frac{\kappa}{4\pi}V^2\left[2 \sqrt{\frac{\kappa}{4\pi}}r_0Q
-\frac{\kappa}{4\pi}Q^2-r_0^2\right]\quad ,
\end{equation}
\begin{equation}
\label{bbb}
\nu''+(\nu')^2=\frac{\kappa}{8\pi}e^{-2 \nu}V'^2\quad .
\end{equation}
The difference of (\ref{aaa1}) and (\ref{bbb}) leads to
\begin{equation}
\label{ccc}
r_0=\sqrt{\frac{\kappa}{4\pi}}Q\quad .
\end{equation}
With (\ref{ccc}), the last of the 4D Einstein-type equation reads
\begin{equation}
\label{ddd}
a^2-\frac{\kappa}{8\pi}Q^2=0\quad .
\end{equation}
The 4D Yang-Mills-type equations reduce to
\begin{equation}
\label{eee}
V''-\nu'V'=0
\end{equation}
with the first integral
\begin{equation}
\label{fff}
V'=\frac{q}{a^2}e^{\nu}
\end{equation}
where $q$ is a constant of integration.
After some reformulations of the field equation
$\tilde{R}_4^4+\tilde{R}_5^5+\tilde{R}_6^6 =0$
(which is reminiscent of the 4D Klein-Gordon-type equation),  we obtain
\begin{equation}
\label{ggg}
\frac{\kappa}{8\pi}(q^2+Q^2)=2a^2 \rightarrow q^2 = Q^2
\end{equation}
Finally, again a special solution of Eq. (\ref{bbb}) with
the first integral (\ref{fff}) taken into account is
\begin{equation}
\label{hhh1}
e^{\nu(r)}=\cosh\left(\frac{r}{a}\right)\quad .
\end{equation}
Using this expression for $e^{\nu (r)}$ it is again
possible to integrate $V'$ of Eq. (\ref{fff}) to obtain
$V(r)$ of a form similar to $\omega (r)$ of Eq. (\ref{dd2})
This set of equations and the given solution exhibits
many common points with the equations and solutions
of the 5D Kaluza-Klein theory and with the SU(2) EYM theory.
There are some differences. For example in comparing the
5D and 7D systems one finds for the latter case that
(using Eqs. (\ref{ccc}) (\ref{ddd}) ) there is a relationship between
the length scale of the inner and outer space, namely
\begin{equation}
\label{40a}
r_0=\sqrt{2}a\quad .
\end{equation}
An interesting feature of the SU(2) EYM system
in particular is that the solutions with their dual
relationship between $Q$ and $g$, their ``magnetic
charge'', and their flux tube geometry
mimic some of the main features of the confinement
mechanism for Yang-Mills theory alone. Thus it might
be conjectured that gravity plays some interesting
role in the confinement mechanism. At first
this conjecture appears completely unreasonable
given the weak strength of the gravitational coupling
compared to the gauge interaction couplings of the
Standard Model. However, other recent works have
explored similar ideas. First, in the large extra
dimension approach \cite{hamed} gravity becomes
strongly coupled at an energy scale much lower
than $10^{19}$ GeV, thus making the gravitational
interaction important at these lower energy scales.
Usually, however, these energy scales are
greater than 1 TeV rather than the 1 GeV energy scale
associated with the strong interaction. Second, an
interesting connection between the glueball mass spectrum
of pure Yang-Mills theory and supergravity was studied in
Refs. \cite{glue}.

\section{Concluding remarks}

The flux tube solutions studied here are significantly
different (especially topologically) from the more commonly
studied spherically symmetric solutions. For
the flux tube solutions, the area of
the 2-spheres ($r=const$) and the 2-disk does not increase
with increasing ``radial coordinate'' $r$. Radial coordinate
is put in quotes since $r$ is not a true radial coordinate
since it runs from $-\infty$ to $+\infty$.
\par
For simplicity, we only considered very special
forms for the metric component $g_{00}$. Other choices
of integration constants would lead to different
behavior of $g_{00}$ for $r\to \infty$, for example, $g_{00}
\to 0$ for $r \rightarrow \pm \infty$. There are also possible
solutions which have a finite upper or lower bound for r.
One could try to ``sew'' together such different
solutions by matching them at some surface, $r=const$.
All these various possible solutions have the common feature
that they are ``cosmological'' solutions - {\it i.e.} solutions
that do not approach a Minkowski form. One way to make
these solutions more physically useful is to view them as
small scale fluctuations of some kind of space-time foam
model. One easy extension of the above work would be to
allow $A^{\bar a}_0$ to have imaginary components
as is sometimes done in pure Yang-Mills gauge
theories \cite{Wu-Yang}. This would change the sign in front of the
third term in equations like Eq. (\ref{14}) and,
the hyperbolic functions would be replaced by trigonometric
functions with a finite range of $r$.
\par
It is remarkable that different types of
gravity theories admit such similar solutions. The
main question concerning these flux tube solutions is
their physical meaning. Maybe, they describe the
ground state of certain physical systems (perhaps the
interior of infinitely extended strings). Apart from
the question of their physical usefulness, if any, these
solutions are interesting from the topological point of view.
\par
Finally, we summarize the following peculiarities
of these solutions :
\begin{itemize}
\item
All these solutions are functionally similar despite coming from
somewhat different theories (4D Einstein gravity plus
either Abelian or non-Abelian gauge fields, or Kaluza-Klein
gravity). The one difference is that
the Kaluza-Klein solutions fix the relative magnitudes of the
electric and magnetic parts of the solutions to be
equal to one another ({\it i.e.} in the 5D Kaluza-Klein
case we found $q=Q$ while in the 4D Einstein-Maxwell
system the magnetic and electric fields could take on any
value with respect to one another). This is similar to the
equality between electric and magnetic parts of the Kaluza-Klein
dyons of Ref. \cite{Perry}.
\item
These gravitational solutions deserve the label
\textit{flux tube spacetimes} since the sections $t=const$,
$r=const'$ have constant area.
\item
These gravitational solutions bear some resemblance to similar
flux tube structures which occur in type-II superconductors and
in QCD. The SU(2) EYM solutions further have a dual
relationship between the ``magnetic'' charge and gauge
coupling, which leads to a confining/deconfining
phase structure with many similarities to the standard
confinement mechanism of pure Yang-Mills theory.
\item
In both the 4D Einstein-Yang-Mills and 7D gravity cases
the gauge fields were only Abelian in nature.
({\it i.e.} even in these non-Abelian cases the fields in the
flux tube were essentially ``electromagnetic'' fields.)
This indicates some type of dynamical reduction of the gauge group :
$SU(2) \rightarrow U(1)$ in the flux tube. This leads to
the possible hypothesis that any gravitational flux
tube could \textit{only} be filled with a U(1) gauge field.
\item
These properties of the flux tube spacetimes
can be summarized graphically:
$$
\begin{array}{ccc}
\biggl (4D \; Einstein-Yang-Mills \biggr ) & \longleftrightarrow &
\biggl (7D \; gravity \biggr )\\
\downarrow & & \downarrow \\
\biggl (4D \; Einstein-Maxwell \biggr ) & \longleftrightarrow &
\biggl (5D \; Kaluza-Klein \biggr )
\end{array}
$$
The vertical arrows indicate a dynamical reduction
$SU(2) \rightarrow U(1)$ and the horizontal arrows indicate a
similarity of the corresponding solutions.
\end{itemize}

\section{Acknowledgments}

VD is grateful for financial support by a Georg Forster Research Fellowship
from the Alexander von Humboldt Foundation and H.-J. Schmidt
for invitation to Potsdam University.

\end{document}